20 DE JUNIO DE 2019

# LOS PERFILES DE INVESTIGACIÓN Y SU IMPLANTACIÓN EN LA UNIVERSIDAD PÚBLICA DE NAVARRA
### INFORME

Manuel Ruiz de Luzuriaga Peña

Isabel Muñoz Mouriño

Mercedes Bogino Larrambebere

Oficina de Referencia

Biblioteca de la Universidad Pública de Navarra

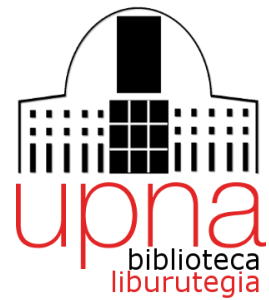

# Sumario

# 1 Introducción

Se conoce como *perfil de investigación* la exposición de datos personales, laborales y de producción científica del personal investigador, realizado por determinadas plataformas, herramientas y servicios.

Las plataformas donde aparecen los perfiles de investigación suelen ofrecerlos como complemento o añadido de otros servicios. La función principal de estas plataformas puede ser como repositorio, red social científica o apoyo de base de datos.

Estos perfiles tienen una serie de ventajas para el personal investigador:

- Presenta agrupada toda la producción científica, a modo de currículum investigador.
- Los perfiles suelen proporcionar un código que sirve como identificación, y que puede ser solicitado por las revistas en las que se publique o en convocatorias o procesos de evaluación.
- Facilita la normalización del nombre, agrupando las publicaciones con independencia de las variantes de nombre utilizadas o de las instituciones en las que se haya trabajado y diferenciándolo también de otros investigadores con igual o similar nombre.
- Potencia la visibilidad y difusión de la producción científica.
- En muchas ocasiones, facilitan datos bibliométricos sobre el investigador: número de citas a sus trabajos, índice h, etc.

Por todo lo anterior, es interesante que investigadores e investigadoras estén presentes en los servicios que ofrecen estos perfiles y sería conveniente que la institución alentara este comportamiento. En la UPNA, a lo largo de 2017, se desarrollaron las siguientes iniciativas por parte de la biblioteca:

- Creación de un grupo de trabajo para identificar duplicados ScopusID del personal investigador de la UPNA y asesorarles sobre su normalización.
- Taller de formación sobre unificación y normalización de identificadores ScopusID.
- Taller de creación de perfiles en ORCID
- Talleres de formación de uso de Mendeley
- Elaboración de guías temáticas sobre [perfiles de investigación en general](#), sobre [ORCID](#), sobre [ResearcherID](#) y sobre [Google Scholar Citations](#).

A lo largo de 2018 se han impartido bibliopartículas (sesiones de formación de una hora) sobre creación y/o unificación de perfiles en Google Scholar Citation, Researcher ID, Scopus ID y ORCID. Además se ha seguido controlando a los investigadores de la UPNA que disponen de perfil en Google Scholar Citation, Researcher ID, Scopus ID y/u ORCID para incluirlo en el [Portal de Producción Científica](#) de la UPNA.

Este trabajo pretende monitorizar y controlar la presencia del personal investigador de la UPNA en las principales plataformas de perfiles de investigación, no solo en las más evidentes como Google Scholar Citation, Researcher ID, Scopus ID y ORCID, sino también en otros servicios que, en la práctica, funcionan como perfiles de investigación, como Mendeley, Linkedin, ResearchGate, Academia.edu y Academica-e. También nos parece interesante analizar esa presencia y ver cómo responde a una serie de variables, como son el departamento, el género, la categoría laboral, el grupo de investigación, etc.

Una tipología de las plataformas que suministran servicios de perfil de investigación podría ser la siguiente:

1. **Perfil de investigación institucional:**
    - [PPC Portal de Producción Científica de la UPNA](#)
2. **Perfiles de investigación en sentido estricto:**
    - [ORCID](#)
3. **Perfiles de investigación ligados a una base de datos:**
    - [ResearcherID (Web of Science)](#)[1]
    - [ScopusID (Scopus)](#)
    - [Google Scholar Citations (Google Scholar)](#)
    - [Dialnet (Dialnet)](#)
4. **Perfiles de investigación ligados a un repositorio institucional:**
    - [Academica-e (Repositorio institucional de la UPNA)](#)
5. **Perfiles de investigación ligados a redes sociales científicas/repositorios privados:**
    - [ResearchGate](#)
    - [Academia.edu](#)
6. **Otros servicios que incluyen perfiles de investigación:**
    - [Mendeley (gestor bibliográfico)](#)
    - [Publons (servicio de control de revisiones)](#)[1]
    - [Linkedin (red social de negocios y empresa)](#)

En este estudio hemos excluido algunas plataformas por diferentes motivos. Los perfiles de Dialnet se introducen desde la biblioteca de la UPNA (BUPNA), lo que quiere decir que estarían todos los que cumplen los requisitos para su inclusión, por lo cual no tiene mucho sentido su análisis, ya que depende de factores ajenos a la voluntad del propio investigador. Otro tanto sucede con el Portal de Producción Científica de la UPNA (PPC): los datos se introducen desde el Vicerrectorado de Investigación y deberían estar todos los miembros del PDI de la UPNA. También hemos excluido Publons[1], a pesar de su interés, por el reducido número de investigadores UPNA presentes (en torno a 50), que no permite sacar ninguna conclusión ni obtener ningún patrón.

---

[1] En marzo de 2019 se anuncia la fusión de ResearcherID con la plataforma Publons.

# 2 METODOLOGÍA

Usando como base el censo de personal investigador de la universidad suministrado por el Vicerrectorado de Investigación, se ha comprobado, para cada autor o autora, la existencia o no de un perfil en los diferentes servicios estudiados. Los resultados se han ido tabulando en un fichero Excel para poder analizarlos posteriormente.

Los datos han sido tomados en marzo de 2018 para Orcid, ResearcherID, ScopusID, Google Scholar Citations y Mendeley. En Noviembre de 2018 se tomaron los datos de Academica-e, Academia.edu, ResearchGate y Linkedin.

Para cada uno de los perfiles, se ha utilizado, cuando era posible, una búsqueda por la adscripción institucional, para disponer de un primer listado de personal investigador de la UPNA con ese perfil. Posteriormente, se realizaba una búsqueda, persona por persona, del resto del personal investigador que no aparecía en ese primer listado.

Lógicamente ha habido muchas dudas y problemas: personas con nombre y apellidos muy comunes, falta de datos en el perfil, errores de tecleo, nombres mal puestos, etc. Cuando figura en el perfil la producción bibliográfica, se puede comprobar si es la persona que estamos buscando controlando sus publicaciones. Si no se ha podido verificar de alguna manera, hemos seguido el criterio de considerar únicamente los casos en los que tenemos una certeza razonable de que el perfil que nos aparece se trata, efectivamente, del investigador o investigadora de la UPNA. Esta certeza consiste en la coincidencia de nombre y un apellido o de los dos apellidos si son poco corrientes o, en caso de nombre y apellidos corrientes, de nombre y dos apellidos, y que no haya concurrencia de otro investigador con el mismo nombre y apellidos. En caso de duda, no lo consideramos.

Probablemente, pues, habrá un número algo mayor de personal investigador de la UPNA con perfil en alguno de los servicios estudiados pero que, ante la falta de certeza, no se ha contabilizado. No obstante, consideramos que el peso de estos casos en el total es despreciable a efectos de este informe.

# 3 PERFILES DE INVESTIGACIÓN

## 3.1 ORCID

Open Researcher and Contributor ID (Orcid) fue creado en 2012 (Taylor, 2012) como una organización internacional, sin ánimo de lucro, mantenida por las aportaciones de sus socios, que van desde universidades e institutos de investigación a editores y asociaciones profesionales, entre otros actores en el ámbito de la investigación.

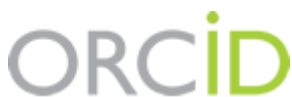

Orcid proporciona un identificador digital único y persistente a cada investigador, que le permite distinguirse de los demás investigadores y evitar errores de autoría por haber firmado su producción científica de múltiples formas. Se compone de 16 números divididos en 4 bloques. Este sería un ejemplo:

0000-0002-1279-6195

Se puede acceder al perfil correspondiente construyendo una URL con el prefijo https://orcid.org/ y el identificador Orcid.

https://orcid.org/0000-0002-1279-6195

Cuando un investigador se registra, puede incorporar los datos bibliográficos de su producción científica, un currículum académico y laboral e incluso los organismos financiadores de su investigación. Además, permite relacionar el código Orcid con otros identificadores como son el Scopus Author ID, ResearcherID o el ISNI.

Según la página web de Orcid, actualmente hay asignados 5.957.040 identificadores Orcid, y, sin duda, este número irá en aumento, debido a que es un servicio ampliamente aceptado por la comunidad investigadora e incluso es un requisito a la hora de publicar o solicitar financiación (Martín, 2019).

Orcid tiene la ventaja de que no está subordinado y es independiente de cualquier editor o plataforma comercial, por lo que puede hacer de puente entre los distintos códigos e identificadores proporcionados por una plataforma concreta.

**En la UPNA**, de un total de 1081 investigadores, 479 tienen perfil en Orcid, lo que supone un 44,31% del total.

De esos 479 perfiles, solamente 297 han introducido algún dato bibliográfico o un enlace a otro identificador, es decir un 62%. El 38% restante de los perfiles Orcid de investigadores de la UPNA está vacío, suponemos que por desconocimiento del proceso o, que una vez obtenido el código que se necesita para algún trámite administrativo, no ha habido interés en completar el perfil. Esos 297 perfiles con datos suponen un 27,47% del total de 1081 investigadores.

A pesar de las campañas de promoción, el uso de Orcid sigue siendo relativamente bajo. Más, si tenemos en cuenta que, del 44,31% de ese total de investigadores UPNA con perfil Orcid, solo un 27,47% tiene lo completado mínimamente.

Apenas hay **estudios sobre el uso de Orcid** entre los investigadores que nos permitan comparar este resultado. En un estudio sobre las universidades gallegas realizado mediante encuesta (Rodríguez Fernández et al., 2018) se indica una adscripción a Orcid de 81,19%[2]. Otro trabajo también mediante encuesta indica un 15% de investigadores usando ORCID en la Stony Brook University (Tran y Lyon, 2017). En la Universidad de Leiden se menciona un 68,9% de investigadores de la universidad con perfil en Orcid (Verhaar, 2018). Como se ve, es difícil la comparación, tanto por el ámbito geográfico como por el método seguido, ya que dos de los tres trabajos se realizan mediante encuesta.

En una **distribución por departamentos**, observamos una variación desde el 65,52% de los investigadores del Departamento de Gestión de Empresas al 27,78% del Departamento de Derecho. No hay una distinción clara entre los departamentos de ciencia y tecnología y los de humanidades y ciencias sociales: Sociología y Trabajo Social está bastante arriba y Ciencias de la Salud muy abajo. Aparte de la propia materia, también influye en gran medida la composición interna del departamento, la situación contractual, como veremos más adelante.

| Departamento | n | % de ORCID | % ORCID+ |
|---|---|---|---|
| GESTIÓN DE EMPRESAS | 58 | 65,52% | 36,21% |
| INGENIERÍA ELÉCTRICA, ELECTRÓNICA Y DE COM | 137 | 56,20% | 36,50% |
| ECONOMÍA | 58 | 51,72% | 27,59% |
| CIENCIAS | 96 | 51,04% | 40,63% |
| ESTADÍSTICA, INFORMÁTICA Y MATEMÁTICAS | 121 | 50,41% | 36,36% |
| SOCIOLOGÍA Y TRABAJO SOCIAL | 92 | 47,83% | 21,74% |
| AGRONOMÍA, BIOTECNOLOGÍA Y ALIMENTACIÓN | 76 | 44,74% | 28,95% |
| INSTITUTO AGROBIOTECNOLOGIA | 25 | 40,00% | 32,00% |
| INGENIERÍA | 121 | 36,36% | 25,62% |
| CIENCIAS HUMANAS Y DE LA EDUCACIÒN | 86 | 32,56% | 17,44% |
| CIENCIAS DE LA SALUD | 139 | 31,65% | 15,83% |
| DERECHO | 72 | 27,78% | 12,50% |
| **Total general** | **1081** | **44,31%** | **27,47%** |

*Tabla 1 - % de ORCID por departamentos*

También podemos observar en la tabla 1 el porcentaje de investigadores en cada departamento con el perfil de Orcid relativamente completado (% ORCID +)[3]. Aunque existe una correlación importante respecto al total de perfiles de ORCID, hay una pequeña alteración del orden: por ejemplo, el Departamento de Ciencias (40,63%) es el que cuenta con un porcentaje mayor de perfiles completados, seguido por Ingeniería Eléctrica (36,5) y Estadística (36,36%). Los departamentos de menor porcentaje son, como en la clasificación general, Ciencias de la Salud (15,83%) y Derecho (12,5%).

| Situación | n | % de ORCID | % ORCID+ |
|---|---|---|---|
| Ayudante doctor | 54 | 77,78% | 40,74% |
| Contratado doctor | 126 | 72,22% | 51,59% |
| Catedrático | 83 | 63,86% | 49,40% |
| Investigador doctor | 40 | 62,50% | 50,00% |
| Titular | 246 | 61,79% | 41,87% |
| Becario | 110 | 36,36% | 14,55% |
| Asociado | 155 | 28,39% | 11,61% |
| Emérito | 16 | 18,75% | 6,25% |
| Investigador | 82 | 14,63% | 9,76% |
| Doctorando | 169 | 10,06% | 1,78% |
| **Total** | **1081** | **44,31%** | **27,47%** |

*Tabla 2 - % de Orcid por situación contractual*

Si observamos la **distribución por situación contractual**, vemos una variación desde el 77,78 % entre los ayudantes doctores hasta el 10,06% de los doctorandos matriculados en tutela académica. Observamos un corte notable entre los grupos[4] con mayores necesidades de publicación (ayudante doctor, contratado doctor, catedrático, investigador doctor y titular) y el resto. El porcentaje de investigadores con el perfil Orcid relativamente

---

[2] Este estudio se ha realizado a través de una encuesta enviada a todo el PDI de las universidades de Galicia. Solo ha respondido el 8,4%. Hay que suponer un cierto sesgo y que los que responden son precisamente los que más participan en redes sociales científicas.

[3] Consideramos aquí los perfiles que han introducido algún dato bibliográfico o, al menos, un enlace a otro identificador.

[4] Becario engloba todos los contratos predoctorales; Asociado incluye: asociado 1, asociado 2 y asociado 3; Investigador comprende: ayudante de proyecto y colaborador de proyecto; Investigador doctor engloba: colaborador doctor de proyecto y otros contratos postdoctorales.

completado varía muy poco de lo que marca el tenor general de la disposición de perfil Orcid.

| Grupos | n | % de ORCID | % ORCID+ |
|---|---|---|---|
| **Ciencias humanas y sociales** | | | |
| Sociología rural, movilidad e investigación social | 11 | 81,82% | 45,45% |
| Psicología Clínica y Psicopatología | 11 | 63,64% | 45,45% |
| ALTER. Grupo de investigación | 19 | 52,63% | 21,05% |
| Cambios sociales | 20 | 50,00% | 20,00% |
| Evaluación y desarrollo de la percepción musical | 10 | 50,00% | 20,00% |
| Hizkuntzalaritzako ikerketa/Investigación en lingüística | 16 | 31,25% | 12,50% |
| Cultura y Desarrollo Lera-ikergunea | 11 | 9,09% | 0,00% |
| **Ingeniería y tecnología** | | | |
| Comunicación, señales y microondas | 16 | 75,00% | 31,25% |
| Producción animal y calidad y tecnología de la carne | 13 | 69,23% | 30,77% |
| Grupo de Antenas | 14 | 64,29% | 50,00% |
| Ingeniería biomédica | 15 | 60,00% | 33,33% |
| Comunicaciones ópticas y aplicaciones electrónicas | 55 | 56,36% | 36,36% |
| Proyectos, ingeniería rural y energías renovables | 13 | 53,85% | 38,46% |
| THERRAE: Teledetección, Hidrología, Erosión, Riegos y Análisis Estructural. | 21 | 52,38% | 28,57% |
| Ingeniería Eléctrica, Electrónica de Potencia y Energías Renovables (INGEPER) | 23 | 52,17% | 43,48% |
| Agrobiotecnología | 30 | 46,67% | 36,67% |
| Protección de cultivos (PC) | 22 | 40,91% | 31,82% |
| Diseño industrial | 10 | 40,00% | 30,00% |
| Ingeniería Térmica y de Fluidos | 15 | 26,67% | 20,00% |
| Ingeniería de materiales y fabricación | 21 | 23,81% | 14,29% |
| Ingeniería mecánica aplicada y computacional -I.M.A.C.- | 15 | 20,00% | 13,33% |
| Tecnología, control y seguridad alimentaria (ALITEC) | 13 | 7,69% | 0,00% |
| **Ciencias económicas y jurídicas** | | | |
| Economía de la Empresa | 13 | 76,92% | 46,15% |
| Grupo marketing | 12 | 66,67% | 25,00% |
| Organización de empresas | 16 | 56,25% | 37,50% |
| Análisis Económico | 11 | 45,45% | 27,27% |
| Historia y Economía | 26 | 42,31% | 30,77% |
| Economía de la salud | 15 | 40,00% | 13,33% |
| Administración Pública | 11 | 36,36% | 36,36% |
| Hugo Grocio | 18 | 33,33% | 5,56% |
| Derecho privado | 16 | 31,25% | 18,75% |
| Estudios de Derecho Público sobre inmigración, extranjería y multiculturalidad | 10 | 20,00% | 10,00% |
| **Ciencias básicas y de la salud** | | | |
| Inteligencia artificial y razonamiento aproximado | 20 | 60,00% | 45,00% |
| DECYL (Datos, Estadística, Calidad y Logística) | 19 | 57,89% | 42,11% |
| Propiedades físicas y aplicaciones de materiales | 11 | 54,55% | 45,45% |
| Fisiología vegetal y Agrobiología | 22 | 45,45% | 40,91% |
| Grupo de investigación en Saberes Enfermeros (GISE) | 21 | 42,86% | 14,29% |
| Control de la expresión génica | 12 | 41,67% | 16,67% |
| Reactores Químicos y Procesos para la Valorización de Recursos Renovables | 13 | 38,46% | 38,46% |
| Sistemas distribuidos | 24 | 37,50% | 33,33% |
| Estadística espacial | 11 | 36,36% | 18,18% |
| Álgebra. Aplicaciones. | 12 | 33,33% | 25,00% |
| Epidemiología | 15 | 33,33% | 20,00% |
| Biofilms microbianos | 12 | 25,00% | 25,00% |
| Ejercicio Físico, Salud y Calidad de vida (E-FIT) | 30 | 16,67% | 10,00% |
| Fisiopatología y práctica clínica | 18 | 16,67% | 0,00% |
| **Total general** | **782** | **43,99%** | **27,24%** |

*Tabla 3 - % de Orcid por grupo de investigación*

En la **distribución por grupos de investigación**[5] se aprecian las diferencias ya señaladas para los departamentos, con algunas particularidades. Los grupos con mayor presencia de investigadores con Orcid son *Sociología rural, movilidad e investigación social*, con un 81,82%, Economía de la empresa, con un 76,92% y Comunicación, señales y microondas con un 75%. Los grupos de investigación con menos investigadores con Orcid son *Tecnología, control y seguridad alimentaria* (7,69%), *Cultura y desarrollo Lera ikergunea* (9,09%), *Ejercicio físico, salud y calidad de vida* (16,67%) y *Fisiopatología y práctica clínica* (16,67%).

---

[5] Se ha considerado solamente los grupos de investigación con diez o más componentes, por considerar de poco valor estadístico los grupos por debajo de este umbral.

Las áreas en las que se clasifican los grupos de investigación[6] presentan poca variación entre ellas, en un abanico que va del 47,06% del área de Ingeniería y tecnología al 40,6% de Ciencias básicas y de la salud. No se puede aventurar que los investigadores de un área sean más proclives a obtener el Orcid. Parece más significativa la distribución por departamentos o por situación contractual. Predomina la individualidad de los grupos de investigación, con comportamientos distintos respecto a Orcid, incluso dentro de la misma área de conocimiento.

| Áreas | n | % de ORCID | % ORCID+ |
|---|---|---|---|
| Ingeniería y tecnología | 357 | 47,06% | 31,65% |
| Ciencias económicas y jurídicas | 205 | 45,37% | 24,39% |
| Ciencias humanas y sociales | 184 | 44,57% | 21,74% |
| Ciencias básicas y de la salud | 335 | 40,60% | 28,06% |
| **Total** | **1081** | **44,31%** | **27,47%** |

*Tabla 4 - % de Orcid por área de conocimiento*

---

[6] Para hacer este cálculo hemos tomado todos los datos, también los de los grupos con menos de diez investigadores.

## 3.2 ResearcherID[7]

ResearcherID es un identificador digital único de investigador que creó Thomson Reuters en 2008 (Qiu, 2008), en principio, para los autores con trabajos en la Web of Science (WOS), aunque este no es un requisito imprescindible y se pueden incluir autores y trabajos que no estén en la WOS. Según su página web, más de 1 millón de usuarios disponen de este identificador.

Los investigadores tienen que registrarse en la web de ResearcherID y esta les proporciona un identificador compuesto de una letra y 2 bloques de 4 números. El último bloque es el año de creación. Un ejemplo sería el siguiente: H-4868-2011

Los autores pueden rellenar su perfil con datos de afiliación, correo electrónico, páginas web personales, ámbito de investigación, palabras clave sobre sus campos de interés y su identificador ORCID con el que puede intercambiar datos. Estos perfiles pueden ser públicos o privados. Bajo este identificador los autores pueden agrupar todas sus publicaciones que aparezcan en WOS y así evitar problemas de ambigüedades en la autoría, y hacer un seguimiento de sus citas. Researcher ID proporciona el índice-h de autor, el número de citas recibidas, su distribución por años y la media de cita por artículos.

**En la UPNA**, de un total de 1081 investigadores, 247 tienen perfil en ResearcherID, lo que supone un 22,85% del total. Este porcentaje parece un poco escaso, dada la importancia que en algunos ámbitos tiene este perfil[8].

| Departamento | n | % en ResearcherID |
|---|---|---|
| GESTIÓN DE EMPRESAS | 58 | 53,45% |
| CIENCIAS | 96 | 36,46% |
| AGRONOMÍA, BIOTECNOLOGÍA Y ALIMENTACIÓN | 76 | 32,89% |
| ESTADÍSTICA, INFORMÁTICA Y MATEMÁTICAS | 121 | 32,23% |
| INGENIERÍA ELÉCTRICA, ELECTRÓNICA Y DE COM | 137 | 28,47% |
| ECONOMÍA | 58 | 27,59% |
| INGENIERÍA | 121 | 14,88% |
| CIENCIAS DE LA SALUD | 139 | 14,39% |
| SOCIOLOGÍA Y TRABAJO SOCIAL | 92 | 13,04% |
| CIENCIAS HUMANAS Y DE LA EDUCACIÓN | 86 | 9,30% |
| DERECHO | 72 | 5,56% |
| INSTITUTO AGROBIOTECNOLOGIA | 25 | 0,00% |
| **Total general** | **1081** | **22,85%** |

*Tabla 5 - % de ResearcherID por departamentos*

| Situación | n | % en ResearcherID |
|---|---|---|
| Catedrático | 83 | 45,78% |
| Contratado doctor | 126 | 44,44% |
| Titular | 246 | 38,62% |
| Ayudante doctor | 54 | 31,48% |
| Investigador doctor | 40 | 17,50% |
| Becario | 110 | 12,73% |
| Investigador | 82 | 8,54% |
| Emérito | 16 | 6,25% |
| Asociado | 155 | 5,81% |
| Doctorando | 169 | 1,78% |
| **Total general** | **1081** | **22,85%** |

*Tabla 6 - % de ResearcherID por situación contractual*

La **distribución por departamentos** muestra una variación desde el 53,45% del Departamento de Gestión de Empresas al 5,56% de Derecho o a la ausencia de perfiles ResearcherID en el Instituto de Agrobiotecnología. Salvo el caso de Gestión de Empresas y Agrobiotecnología, parece haber una clara diferencia entre los departamentos de ciencia y tecnología y los de ciencias sociales y humanas. Esto parece lógico si atendemos a su carácter semidependiente de la WOS, que hace que parezca lógico que la materialización de los perfiles ResearcherID se corresponda con el contenido de la WOS.

En la **distribución por situación contractual** vemos una situación parecida a la observada para Orcid: tienen mayor porcentaje aquellos grupos más dedicados a la investigación o que demuestran una carrera investigadora más prolongada. Así, tenemos a los catedráticos con un 45,78% y, en el otro extremo, a los doctorandos con un 1,78%.

---

[7] El 15 de abril de 2019 ResearcherID se integrará en la plataforma Publons.
[8] La inclusión de ResearcherID en Publons hace suponer que se apuesta fuerte por esta plataforma que, además, ofrecerá servicios como el control de revisiones, lo cual hace todavía más llamativa la baja inscripción.

## 3.3 SCOPUS AUTHOR ID

Scopus es una base de datos propiedad de Elsevier, desde 2006 (Fenner, 2011). Asigna automáticamente a los autores recogidos un identificador único, bajo el que se agrupan sus trabajos (Martín, 2019). Un ejemplo sería el siguiente:

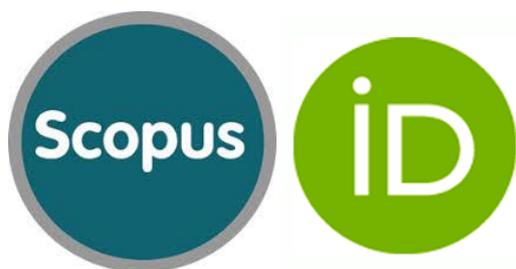

Author ID: 55034746100

Para evitar la homonimia y las diferentes variantes del nombre de un mismo autor, para relacionar un trabajo con un autor, utiliza un algoritmo que tiene en cuenta la afiliación, ciudad, la materia de investigación, fechas de publicación, citas y coautores. Aunque el autor no puede modificar directamente estos datos, si encuentra algún error, puede solicitar a Scopus que lo corrija por medio del Author Feedback Wizard.

En el perfil de cada investigador aparece su ORCID, otros nombres por los que es conocido, sus áreas de investigación y los coautores. En cuanto a métricas, aporta el índice-h de autor, el número de documentos y de citas, así como su distribución por años.

Es un poco diferente de ResearcherID y Google Scholar Citation en el sentido de que no es voluntario. No obstante, lo incluímos en este informe porque creemos que es una buena referencia para compararlo con los otros perfiles de autor.

**En la UPNA** hay 532 investigadores que disponen de un Scopus Author ID, lo que supone un 49,21% del total del PDI. Este porcentaje es mayor que el de ResearcherID e incluso que el de Orcid. Este predominio suponemos que viene dado por la automatización en la creación del perfil si se tienen documentos indizados en Scopus. De esta manera, la creación del perfil no depende de la iniciativa del investigador.

La **distribución por departamentos** es muy similar a la de ResearcherID, solo que con cifras más amplias. Varía desde el 72,41% del Departamento de Gestión de Empresas al 8,33% del Departamento de Derecho. La presencia de los trabajos de un determinado autor en Scopus es requisito imprescindible para disponer de un Scopus Author ID, esto limita bastante a los investigadores del departamento de Derecho, y también a los de Ciencias Humanas (26,74%), que publican mayoritariamente en revistas que no son recogidas en Scopus.

| Departamentos | n | % en Scopus Author ID |
|---|---|---|
| GESTIÓN DE EMPRESAS | 58 | 72,41% |
| INGENIERÍA ELÉCTRICA, ELECTRÓNICA Y DE COM | 137 | 70,07% |
| ESTADÍSTICA, INFORMÁTICA Y MATEMÁTICAS | 121 | 61,98% |
| AGRONOMÍA, BIOTECNOLOGÍA Y ALIMENTACIÓN | 76 | 61,84% |
| CIENCIAS | 96 | 61,46% |
| ECONOMÍA | 58 | 56,90% |
| INGENIERÍA | 121 | 52,07% |
| INSTITUTO AGROBIOTECNOLOGIA | 25 | 36,00% |
| SOCIOLOGÍA Y TRABAJO SOCIAL | 92 | 35,87% |
| CIENCIAS DE LA SALUD | 139 | 33,09% |
| CIENCIAS HUMANAS Y DE LA EDUCACIÒN | 86 | 26,74% |
| DERECHO | 72 | 8,33% |
| **Total general** | **1081** | **49,21%** |

*Tabla 7 - % de Scopus Author ID por departamentos*

| Situación | n | % en Scopus Author ID |
|---|---|---|
| Catedrático | 83 | 75,90% |
| Titular | 246 | 71,95% |
| Contratado doctor | 126 | 69,05% |
| Ayudante doctor | 54 | 61,11% |
| Investigador doctor | 40 | 52,50% |
| Emérito | 16 | 50,00% |
| Becario | 110 | 40,91% |
| Investigador | 82 | 35,37% |
| Asociado | 155 | 33,55% |
| Doctorando | 169 | 10,06% |
| **Total general** | **1081** | **49,21%** |

*Tabla 8 - % de Scopus Author ID por situación contractual*

La **distribución por situación contractual** es relativamente similar a la observada en el resto de perfiles de autor, con valores más elevados para aquellas situaciones contractuales más estables y más vinculadas a la investigación.

## 3.4 GOOGLE SCHOLAR CITATIONS

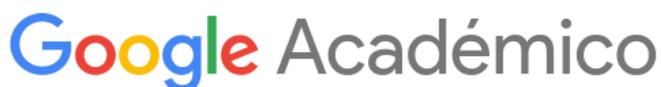

Google Scholar Citations (GSC) ofrece a los investigadores la posibilidad de disponer de un perfil en el que poder unificar sus trabajos y hacer un seguimiento de sus citas. GSC está disponible desde 2011 (Cabezas Clavijo y Torres Salinas, 2012). La ventaja de esta herramienta, sobre todo para investigadores de ciencias sociales y humanidades, es que localiza las citas, no solamente de artículos, sino también de libros, ponencias de congresos, informes, tesis y trabajos depositados en repositorios.

Una vez que el autor se da de alta puede añadir su afiliación, materias de investigación, reclamar trabajos para su bibliografía, eliminar los que el buscador le haya asignado de manera errónea y añadir coautores.

La bibliografía del autor aparece ordenada por el número de citas recibidas. Con estas citas el sistema genera las siguientes métricas: número de citas totales, índice-h e índice-i10 y un gráfico con la distribución de citas por año.

El perfil puede ser público o privado y tiene la opción de realizar actualizaciones automáticamente.

GSC se complementa con Publish or Perish (Harzing, 2011), un software que permite la búsqueda y explotación de los perfiles GSC de un modo más amigable.

**En la UPNA** 190 miembros del PDI con perfil GSC, lo que supone un 17,58% del personal docente e investigador de la UPNA. Es un resultado mucho menor que para los otros perfiles de autor que hemos analizado. Consideramos que es una cifra muy baja de adscritos, teniendo en cuenta las ventajas que proporciona un perfil en estos servicios, sobre todo para investigadores de ciencias sociales y humanidades, que tienen menos presencia y control de las citas a sus trabajos en Scopus y Web of Science.

La **distribución por departamentos** se diferencia un poco de los otros perfiles en que los departamentos de Agronomía e Ingeniería ocupan puestos más bajos. La lista está encabezada —como sucede en los otros perfiles— por el departamento de Gestión de Empresas con un 34,48% y cerrada por el departamento de Derecho (6,49%) e Instituto de Agrobiotecnología sin ningún perfil GSC. Se hubiera podido esperar que departamentos como Sociología y Trabajo Social (20,65%), Ciencias Humanas y de la Educación (18,6%) y Derecho (6,94%) tuvieran mayores puntuaciones, dadas las ventajas que ofrece GSC sobre los otros perfiles, al cubrir una cantidad mucho mayor de trabajos, no solo por temática, sino también por tipo de documentos, ya que GSC controla las citas de libros, congresos y otros soportes documentales.

| Departamento | n | % de GSC |
|---|---|---|
| GESTIÓN DE EMPRESAS | 58 | 34,48% |
| INGENIERÍA ELÉCTRICA, ELECTRÓNICA Y DE COM | 137 | 29,20% |
| ESTADÍSTICA, INFORMÁTICA Y MATEMÁTICAS | 121 | 25,62% |
| ECONOMÍA | 58 | 24,14% |
| SOCIOLOGÍA Y TRABAJO SOCIAL | 92 | 20,65% |
| CIENCIAS HUMANAS Y DE LA EDUCACIÒN | 86 | 18,60% |
| CIENCIAS | 96 | 13,54% |
| AGRONOMÍA, BIOTECNOLOGÍA Y ALIMENTACIÓN | 76 | 11,84% |
| CIENCIAS DE LA SALUD | 139 | 9,35% |
| INGENIERÍA | 121 | 8,26% |
| DERECHO | 72 | 6,94% |
| INSTITUTO AGROBIOTECNOLOGIA | 25 | 0,00% |
| **Total general** | **1081** | **17,58%** |

*Tabla 9 - % GSC por departamento*

| Situación | n | % de GSC |
|---|---|---|
| Contratado doctor | 126 | 37,30% |
| Ayudante doctor | 54 | 35,19% |
| Catedrático | 83 | 27,71% |
| Titular | 246 | 23,17% |
| Becario | 110 | 14,55% |
| Asociado | 155 | 9,68% |
| Investigador doctor | 40 | 7,50% |
| Emérito | 16 | 6,25% |
| Investigador | 82 | 6,10% |
| Doctorando | 169 | 2,37% |
| **Total general** | **1081** | **17,58%** |

*Tabla 10 - % GSC por situación contractual*

La **distribución por situación contractual** sigue la misma tendencia que los otros perfiles, ocupando los primeros lugares aquellas situaciones que indican perfiles más avanzados en la carrera investigadora.

## 3.5 ACADEMICA-E

Academica-e es el repositorio institucional de la Universidad Pública de Navarra (UPNA). Fue creado en 2011. Alberga 14.289 objetos digitales entre los que podemos encontrar artículos, libros, tesis doctorales, trabajos fin de estudios, actas de congresos y publicaciones de miembros de la UPNA, la Revista Huarte de San Juan y publicaciones de fondo antiguo. Actualmente, en Academica-e, hay registrados 903 investigadores como autores UPNA.

El repositorio ofrece, como valor añadido, métricas de Almetrics, PlumX, Dimensions, Google Scholar, Scopus y Web of Science (WOS).

A pesar de ser de una naturaleza bastante distinta que los perfiles de autor vistos hasta ahora, creemos que su uso admite una buena comparación con ellos, por lo que el acto de depositar una publicación en el repositorio supone de acto voluntario[9], para dar visibilidad, fijar, preservar y facilitar el acceso a sus trabajos, y cuyos fines últimos no difieren del uso de los perfiles de autor más heterodoxos.

**En la UPNA** hay 485 miembros del PDI que han depositado publicaciones en el repositorio[10], lo que supone un 44,87% del total de la plantilla PDI. El resultado es bastante similar al obtenido para Orcid (44,31%), está por debajo del de Scopus Author ID (49,21%) y muy por encima del de ResarcherID (22,85%) y GSC (17,58%).

El número de PDI de la UPNA que deposita sus trabajos en el repositorio Academica-e (44,87%) solo es mínimamente superior al de los que lo hacen en ResearchGate (44,31%) y algo inferior a la suma de ResearchGate y Academia.edu (50,97%)[11].

La **distribución por departamentos** se muestra un poco diferente a los otros perfiles. Las diferencias más notables serían el segundo puesto de ciencias Humanas y de la Educación (58,14%) y las bajas posiciones de Agronomía (42,11%) y Ciencias (43,75%). La primera posición la ocupa el Departamento de Economía (67,24%) y los últimos lugares son para Ciencias de la Salud (25,18%) y el Instituto de Agrobiotecnología (24%).

| Departamento | n | % de Academica-e |
|---|---|---|
| ECONOMÍA | 58 | 67,24% |
| CIENCIAS HUMANAS Y DE LA EDUCACIÒN | 86 | 58,14% |
| GESTIÓN DE EMPRESAS | 58 | 55,17% |
| INGENIERÍA ELÉCTRICA, ELECTRÓNICA Y DE COM | 137 | 54,01% |
| SOCIOLOGÍA Y TRABAJO SOCIAL | 92 | 48,91% |
| ESTADÍSTICA, INFORMÁTICA Y MATEMÁTICAS | 121 | 47,93% |
| CIENCIAS | 96 | 43,75% |
| AGRONOMÍA, BIOTECNOLOGÍA Y ALIMENTACIÓN | 76 | 42,11% |
| DERECHO | 72 | 40,28% |
| INGENIERÍA | 121 | 35,54% |
| CIENCIAS DE LA SALUD | 139 | 25,18% |
| INSTITUTO AGROBIOTECNOLOGIA | 25 | 24,00% |
| **Total general** | **1081** | **44,87%** |

*Tabla 11 - % Academica-e por departamento*

---

[9] No obstante, hay que tener cautela con las comparaciones, porque un trabajo introducido en el repositorio, no siempre es por voluntad del autor: el personal de la biblioteca introduce sistemáticamente los trabajos en abierto que localizan. Por otra parte, el acuerdo sobre la política institucional de acceso abierto en la UPNA establece explícitamente la obligación, para los investigadores de la UPNA, de depositar en Academica-e las publicaciones de investigaciones financiadas con fondos públicos.

[10] Se han excluído las publicaciones en las que el autor sea director o tutor, como tesis o trabajos de fin de estudios.

[11] Lógicamente, se trata de una suma ponderada, contabilizando los investigadores que tienen cuenta en una u otra plataforma, o en ambas.

| Situación | n | % de Academica-e |
|---|---|---|
| Catedrático | 83 | 80,72% |
| Emérito | 16 | 75,00% |
| Contratado doctor | 126 | 72,22% |
| Titular | 246 | 63,01% |
| Ayudante doctor | 54 | 55,56% |
| Asociado | 155 | 41,94% |
| Investigador doctor | 40 | 37,50% |
| Becario | 110 | 26,36% |
| Investigador | 82 | 20,73% |
| Doctorando | 169 | 2,37% |
| **Total general** | **1081** | **44,87%** |

*Tabla 12 - % Academica-e por situación contractual*

La **distribución por situación contractual** sigue el tenor de los demás perfiles de autor y dominan los que están en la cima de la carrera investigadora. En primer lugar, los catedráticos (80,72%) y en último, los doctorandos (2,37%).

## 3.6 ResearchGate

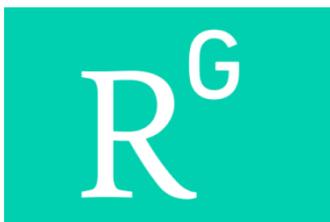

Red social para investigadores fundada en Alemania en 2008 (Orduña-Malea et al., 2016) que cuenta hoy en día con más de 15 millones de usuarios y 118 millones de publicaciones.

ResearchGate permite elaborar un perfil con el currículum, datos personales como dirección de e-mail, teléfono o página web personal, publicaciones y ORCID. Al igual que Academia.edu, permite subir publicaciones a texto completo, lo que lo convierte en un mixto entre red social y repositorio.

Incluye un apartado específico para ponerse en contacto con otros usuarios y comenzar una conversación que puede ser seguida por otros colegas. Otro de sus apartados destacados es el de ofertas de empleo.

En cuanto a métricas, permite ver las citas, recomendaciones y lecturas de las publicaciones. Aparte, proporciona el índice-h de autor y su indicador de reputación RG Score, calculado según un algoritmo que mide la interacción de otros usuarios con las publicaciones. No obstante, esta métrica ha sido criticada con dureza, por medir en mayor medida la actividad en la plataforma que el currículum investigador (Orduña-Malea et al., 2017).

La popularidad de estos servicios de perfil investigador mezcla de red social y repositorio hace que consideremos interesante una comparación con otros servicios de perfiles y con los perfiles de autor en sentido estricto.

**En la UPNA** hay 479 investigadores con cuenta ResearcGate, un 44,31% del total.

En la **distribución por departamentos**, observamos un comportamiento parecido al del resto de los perfiles de investigación: lo único que cambia sería una mayor diferencia entre los primeros y los últimos. Tenemos en la primera posición al Departamento de Gestión de Empresas, con un 87,93% de sus miembros con un perfil ResearchGate y en última al Departamento de Derecho, con solo un 16,67%. La diferencia, como vemos, es notable.

| Departamento | n | % de ResearchGate |
|---|---|---|
| GESTIÓN DE EMPRESAS | 58 | 87,93% |
| INGENIERÍA ELÉCTRICA, ELECTRÓNICA Y DE COM | 137 | 57,66% |
| CIENCIAS | 96 | 55,21% |
| ECONOMÍA | 58 | 55,17% |
| AGRONOMÍA, BIOTECNOLOGÍA Y ALIMENTACIÓN | 76 | 51,32% |
| ESTADÍSTICA, INFORMÁTICA Y MATEMÁTICAS | 121 | 50,41% |
| INGENIERÍA | 121 | 41,32% |
| CIENCIAS HUMANAS Y DE LA EDUCACIÓN | 86 | 31,40% |
| CIENCIAS DE LA SALUD | 139 | 30,94% |
| SOCIOLOGÍA Y TRABAJO SOCIAL | 92 | 29,35% |
| INSTITUTO AGROBIOTECNOLOGIA | 25 | 20,00% |
| DERECHO | 72 | 16,67% |
| **Total general** | **1081** | **44,31%** |

*Tabla 13 - % Researchgate por departamento*

| Situación | n | % de ResearchGate |
|---|---|---|
| Contratado doctor | 126 | 65,87% |
| Ayudante doctor | 54 | 64,81% |
| Titular | 246 | 56,10% |
| Catedrático | 83 | 53,01% |
| Investigador doctor | 40 | 52,50% |
| Becario | 110 | 50,91% |
| Investigador | 82 | 36,59% |
| Asociado | 155 | 33,55% |
| Emérito | 16 | 18,75% |
| Doctorando | 169 | 10,06% |
| **Total general** | **1081** | **44,31%** |

*Tabla 14 - % ResearchGate por situación contractual*

La **distribución por situación contractual** nos muestra una imagen parecida al resto de perfiles, con los investigadores más dedicados y los de carrera más larga en las primeras plazas. Así, el primer lugar lo ocupan los contratados doctores, un 65,8% de los cuales tiene perfil en ResearchGate y la última posición es para los doctorandos, con un 10,06%.

## 3.7 ACADEMIA.EDU

**ACADEMIA** Red social para investigadores creada por Richard Price en 2008 (Niyazov et al., 2016) que, a pesar de su extensión ".edu", pertenece a un fondo de inversión privado (Bond, 2017). Hoy en día, según la propia plataforma, cuenta con más de 73 millones de usuarios y 22 millones de artículos.

Tiene una versión gratuita y otra de pago. La primera permite crear un perfil con tu currículum, establecer tus intereses con palabras clave, compartir las publicaciones, así como ver las veces que se han leído, además de localizar y descargar en texto completo los artículos subidos por otros autores a Academia.edu.

La versión de pago (79 € anuales), además de lo anterior, permite ver las citas que reciben los artículos, datos sobre los lectores de los artículos, la creación de una página web personal y, por último, acceso a convocatorias de becas y subvenciones.

A pesar de que en sus condiciones de uso indica que no se pueden subir publicaciones que infrinjan derechos de autor, editoriales como Elsevier han tenido que solicitar que eliminaran artículos con su copyright (Duffy y Pooley, 2017).

Su estructura es parecida a ResearchGate: la conocida mezcla de perfil de autor, red social y repositorio, con la diferencia que Academia.edu ofrece un servicio Premium para suscriptores, mientras que Researchgate, de momento, sigue siendo gratuito.

**En la UPNA** hay 224 investigadores con perfil en Academia.edu, lo que supone un 20,72% del total. Sale perdiendo en comparación con su rival ResearchGate, en la que, recordemos, había un 44,31% de investigadores UPNA.

En la **distribución por departamentos** observamos que los primeros puestos están ocupados por los departamentos de Sociología y Trabajo Social (39,13%) y Ciencias Humanas y de la Educación (34,88%). Los últimos lugares son para Ingeniería (9,92%) y Agronomía (7,89%). Se deja notar una preferencia por esta red social en humanidades y ciencias sociales.

| Departamentos | n | % de Academia.edu |
|---|---|---|
| SOCIOLOGÍA Y TRABAJO SOCIAL | 92 | 39,13% |
| CIENCIAS HUMANAS Y DE LA EDUCACIÓN | 86 | 34,88% |
| GESTIÓN DE EMPRESAS | 58 | 34,48% |
| INGENIERÍA ELÉCTRICA, ELECTRÓNICA Y DE COM | 137 | 24,82% |
| ECONOMÍA | 58 | 18,97% |
| ESTADÍSTICA, INFORMÁTICA Y MATEMÁTICAS | 121 | 18,18% |
| DERECHO | 72 | 18,06% |
| CIENCIAS | 96 | 16,67% |
| INSTITUTO AGROBIOTECNOLOGIA | 25 | 16,00% |
| CIENCIAS DE LA SALUD | 139 | 14,39% |
| INGENIERÍA | 121 | 9,92% |
| AGRONOMÍA, BIOTECNOLOGÍA Y ALIMENTACIÓN | 76 | 7,89% |
| **Total general** | **1081** | **20,72%** |

*Tabla 15 - % Academia.edu por departamento*

| Situación | n | % de Academia.edu |
|---|---|---|
| Contratado doctor | 126 | 35,71% |
| Ayudante doctor | 54 | 25,93% |
| Catedrático | 83 | 25,30% |
| Emérito | 16 | 25,00% |
| Titular | 246 | 21,95% |
| Asociado | 155 | 20,00% |
| Becario | 110 | 17,27% |
| Investigador | 82 | 14,63% |
| Investigador doctor | 40 | 12,50% |
| Doctorando | 169 | 11,24% |
| **Total general** | **1081** | **20,72%** |

*Tabla 16 - % Academia.edu por situación*

La **distribución por situación contractual** nos muestra una situación similar a la observada para otros perfiles, con la excepción de los investigadores doctores, que caen más bajo de lo habitual. En cabeza están los contratados doctores, con un 35,71% y en la cola, como en otros perfiles, los doctorandos, con un 11,24%.

## 3.8 Mendeley

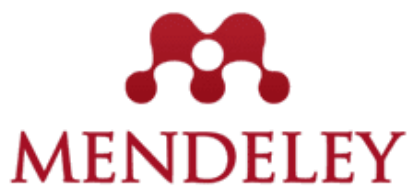

Mendeley fue creado en 2008 por tres estudiantes de informática alemanes recién doctorados que establecieron su sede en Londres. Desde 2013 pertenece a la multinacional de la información Elsevier (Ortega, 2016). Principalmente, es conocido como gestor bibliográfico, pero también es una gran red social científica, con más de 6 millones de usuarios registrados (Martín, 2019). En su faceta de red social, permite crear un perfil en el que el investigador aporta datos académicos y profesionales, las publicaciones realizadas y otros identificadores de perfil de investigación.

Mendeley también permite seguir y ser seguido por otros usuarios, unirse a grupos de investigadores con intereses comunes, localizar literatura especializada, compartir artículos y a la vez promocionar la propia producción científica. Proporciona métricas muy interesantes sobre las publicaciones que un investigador incluye en su perfil: las citas que reciben los trabajos recogidos en Scopus, y los *readers* o veces que una publicación incluida en su perfil aparece en el gestor bibliográfico de otros usuarios (Martín-Martín et al., 2018).

A pesar de ser un poco atípico como perfil, hemos decidido incluir Mendeley en esta comparativa porque tiene muchas similitudes con los perfiles en sentido estricto: la voluntariedad, la recogida de las publicaciones del autor, el suministro de métricas, etc. El problema puede ser que no sabemos si hay un verdadero uso como perfil o el investigador se limita a usarlo como gestor bibliográfico, sin aprovechar los servicios de valor añadido que suministra Mendeley.

**En la UPNA** hay 374 miembros del PDI con cuenta Mendeley, un 34,6% del total. Estas cifras de Mendeley, sin ser ninguna sorpresa ni especialmente dramáticas, indican un uso razonable de este servicio, pero también consideramos que se podrían mejorar. La asistencia a los cursos que se imparten desde la BUPNA[12] indica que hay inquietud por el uso de este gestor bibliográfico y de sus servicios añadidos.

En la **distribución por departamentos** podemos observar algunas diferencias con otros perfiles: el Departamento de Economía ocupa la penúltima posición, con un 25,86% y el Instituto de Agrobiotecnología, la tercera, no siendo estos sus lugares habituales. El primer lugar es para Agronomía, (57,89%), seguido de Ciencias (50%). Las últimas posiciones las ocupan el, ya citado, Departamento de Economía (25,86%) y el de Derecho (13,89%).

| Departamento | n | % de Mendeley |
|---|---|---|
| AGRONOMÍA, BIOTECNOLOGÍA Y ALIMENTACIÓN | 76 | 57,89% |
| CIENCIAS | 96 | 50,00% |
| INSTITUTO AGROBIOTECNOLOGIA | 25 | 44,00% |
| GESTIÓN DE EMPRESAS | 58 | 43,10% |
| INGENIERÍA ELÉCTRICA, ELECTRÓNICA Y DE COM | 137 | 39,42% |
| INGENIERÍA | 121 | 36,36% |
| SOCIOLOGÍA Y TRABAJO SOCIAL | 92 | 30,43% |
| ESTADÍSTICA, INFORMÁTICA Y MATEMÁTICAS | 121 | 28,93% |
| CIENCIAS HUMANAS Y DE LA EDUCACIÒN | 86 | 26,74% |
| CIENCIAS DE LA SALUD | 139 | 26,62% |
| ECONOMÍA | 58 | 25,86% |
| DERECHO | 72 | 13,89% |
| **Total general** | **1081** | **34,60%** |

*Tabla 17 - % Mendeley por departamento*

---

[12] Desde el año 2015, en la biblioteca de la UPNA, se imparten tres talleres al año sobre Mendeley dirigidos al PDI, con unos 25 asistentes en cada uno.

| Situación | n | % de Mendeley |
|---|---|---|
| Becario | 110 | 65,45% |
| Contratado doctor | 126 | 46,03% |
| Ayudante doctor | 54 | 40,74% |
| Investigador | 82 | 35,37% |
| Titular | 246 | 33,33% |
| Investigador doctor | 40 | 32,50% |
| Catedrático | 83 | 28,92% |
| Asociado | 155 | 25,16% |
| Doctorando | 169 | 20,12% |
| Emérito | 16 | 6,25% |
| **Total general** | **1081** | **34,60%** |

*Tabla 18 - % Mendeley por situación contractual*

La **distribución por situación contractual** muestra a los becarios (65,45%) ocupando la primera posición, algo inusual en otros perfiles, donde ocupan posiciones más bajas[13]. La segunda posición, a bastante distancia es para los contratados doctores (46,03%). Los últimos lugares son para los doctorandos (20,12%) y eméritos (6,25%).

---

[13] Una explicación para esta situación podría ser que, en los grupos de investigación, es frecuente que sean los becarios los encargados de la redacción de las citas y bibliografía y tienen mayor necesidad de manejar Mendeley. También suelen ser los becarios los que se forman en Mendeley para, a su vez, formar al resto del grupo o departamento.

## 3.9 LinkedIn

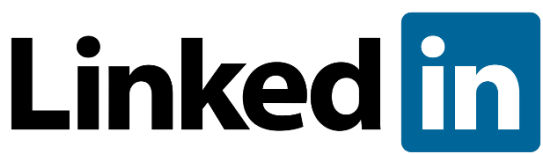

LinkedIn es una red social orientada al mundo laboral que tiene más de 600 millones de usuarios en más de 200 países[14]. Fue creada en 2002 por Reid Hoffman y otros colaboradores (Ollé y López-Borrull, 2017), pero en 2016 pasó a manos de Microsoft.

En el perfil se puede incorporar el currículum académico y laboral, publicaciones, patentes, proyectos, premios e idiomas. Tiene una versión básica que te permite seguir a otros usuarios o grupos, publicar noticias y acceder a ofertas de empleo. La versión *premium* ofrece una mayor visibilidad del perfil ante posibles empleadores, seguimiento de todas las personas que han visto tu perfil y comparación con otros candidatos a un puesto de trabajo. También hay versiones *premium* para empresas, ventas y contrataciones.

A pesar de no ser una red social científica, muchos investigadores tienen presencia en ella. Hay foros y comunidades de corte científico y es una herramienta en la que los medios suelen reclutar expertos sobre una determinada materia. Por todo ello la hemos incluido en esta comparación.

**En la UPNA** hay 536 investigadores con cuenta de Linkedin, 49,58% del total del PDI. Es el perfil que cuenta con el mayor número de adscripciones, por encima del 49,21% del Scopus Author ID o del 44,31% de Orcid. Al ser una red social semigeneralista parte con una ventaja sobre las redes sociales científicas y los perfiles en sentido estricto. Suponemos que el principal atractivo será su utilidad como red social profesional, que no académica.

En la **distribución por departamentos** se aprecia que departamentos que ocupaban las últimas posiciones en los demás perfiles aquí ocupan puestos en la zona media-alta; este es el caso de Derecho, quinta posición con un 56,94% y Ciencias de la Salud, séptima posición con un 49,64%. El primer lugar es para el Departamento de Economía, con un 63,79% y los últimos para Sociología y Trabajo Social (39,13%) y Ciencias Humanas y de la Educación (34,88).

| Departamento | n | % de Linkedin |
|---|---|---|
| ECONOMÍA | 58 | 63,79% |
| INSTITUTO AGROBIOTECNOLOGIA | 25 | 60,00% |
| GESTIÓN DE EMPRESAS | 58 | 58,62% |
| INGENIERÍA ELÉCTRICA, ELECTRÓNICA Y DE COM | 137 | 58,39% |
| DERECHO | 72 | 56,94% |
| INGENIERÍA | 121 | 52,07% |
| CIENCIAS DE LA SALUD | 139 | 49,64% |
| CIENCIAS | 96 | 45,83% |
| AGRONOMÍA, BIOTECNOLOGÍA Y ALIMENTACIÓN | 76 | 44,74% |
| ESTADÍSTICA, INFORMÁTICA Y MATEMÁTICAS | 121 | 43,80% |
| SOCIOLOGÍA Y TRABAJO SOCIAL | 92 | 39,13% |
| CIENCIAS HUMANAS Y DE LA EDUCACIÓN | 86 | 34,88% |
| **Total general** | **1081** | **49,58%** |

*Tabla 19 - % Linkedin por departamento*

| Situación | n | % de Linkedin |
|---|---|---|
| Investigador | 82 | 60,98% |
| Becario | 110 | 58,18% |
| Ayudante doctor | 54 | 57,41% |
| Investigador doctor | 40 | 52,50% |
| Titular | 246 | 50,81% |
| Catedrático | 83 | 49,40% |
| Asociado | 155 | 48,39% |
| Contratado doctor | 126 | 47,62% |
| Emérito | 16 | 43,75% |
| Doctorando | 169 | 36,69% |
| **Total general** | **1081** | **49,58%** |

*Tabla 20 - % Linkedin por situación contractual*

La **distribución por situación contractual** muestra, como en el resto de perfiles, a los doctorandos en última posición (36,69%), pero las primeras posiciones son para investigadores (60,98%) y becarios (58,18%) que, en los demás perfiles —salvo en Mendeley— suelen ocupar las últimas posiciones.

---

[14] Datos tomados de Wikipedia en inglés: https://en.wikipedia.org/wiki/LinkedIn

# 4 ANÁLISIS

## 4.1 PANORAMA GLOBAL

Comparando la presencia global de los miembros del PDI de la UPNA en los distintos servicios de perfiles, podemos observar un panorama en el cual en ningún perfil se supera el 50% de adscripción y en alguno, como es el caso de Google Scholar Citations, no se llega al 20%.

El perfil más solicitado es Linkedin (49,58%), red social semigeneralista y que, quizá, no se relaciona tanto con la investigación como la que figura en segundo lugar, que es Scopus Author ID (49,21%).

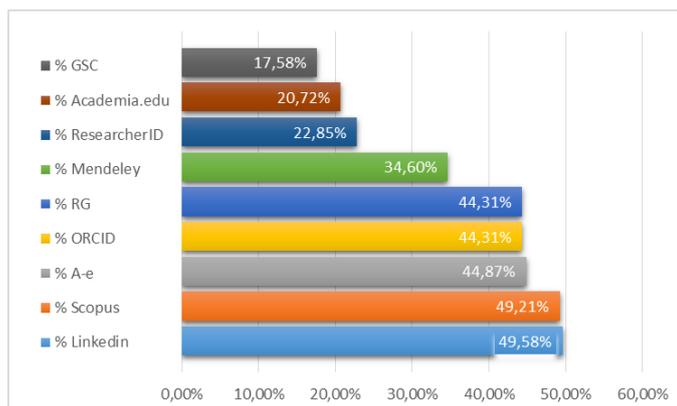

*Gráfico 1 - % global de la UPNA en los diferentes servicios de perfiles*

957 miembros del PDI tienen al menos un perfil de los aquí estudiados (88,53%). Así pues, un 11,47% (124) del PDI no tiene ningún perfil. Lo más repetido es que dispongan de un solo perfil, situación en la que se hallan 191 investigadores (17,67%). Finalmente, tenemos a un 1,85% del PDI (20) que tiene cuenta en los nueve perfiles estudiados.

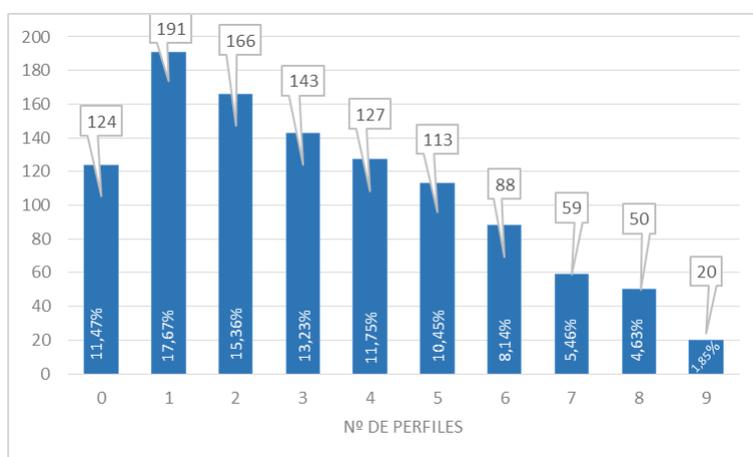

*Gráfico 2 - Distribución por número de perfiles*

## 4.2 Departamentos

En la *Tabla 21* se muestra el cruce comparativo de los departamentos con los distintos perfiles. El color de las celdas, de verde a rojo, marca el valor del dato entre todos los perfiles. Los puntos de color, de verde a negro, marcan el valor dentro del mismo perfil.

El orden de los departamentos lo marca el dato de la media de perfiles por investigador (perfiles/n).

También incluimos el dato de porcentaje de investigadores con al menos un perfil de los estudiados en cada departamento que, aunque coincide en líneas generales con la media de perfiles por investigador, muestra algunas particularidades: Agronomía tiene al 100% de sus 76 investigadores con al menos un perfil, pero ocupa el sexto lugar en número de perfiles por investigador (3,55); El Instituto de Agrobiotecnología ocupa un lugar medio (92%) en investigadores con al menos un perfil, pero es de los más bajos (2,4) en número de perfiles por investigador. El caso contrario son los departamentos de Estadística y Sociología, que están en la zona media considerando el número de perfiles por investigador, pero son de los más bajos en investigadores con al menos un perfil.

Observamos como Gestión de Empresas, Ingeniería Eléctrica y Economía son los departamentos más proclives al uso de los perfiles de investigación, mientras que Derecho, Ciencias de la Salud, Ciencias Humanas, Ingeniería y Sociología, en el otro extremo, son los más reacios a usar los servicios de perfiles.

También se aprecia una preferencia en determinados departamentos por un servicio concreto, con unas cifras que se alejan del comportamiento de ese departamento respecto a otros servicios. Es el caso de Agronomía con Mendeley, Ciencias Humanas con Academica-e y Academia.edu, Derecho con Linkedin, Gestión de Empresas con ResearchGate y Sociología con Academia.edu.

| Departamentos | n | % Linkedin | % Scopus | % A-e | % ORCID | % RG | % Mendeley | % ResercID | % Acade.edu | % GSC | % de >0 | perfiles/n |
|---|---|---|---|---|---|---|---|---|---|---|---|---|
| GESTIÓN DE EMPRESAS | 58 | 58,62% | 72,41% | 55,17% | 65,52% | 87,93% | 43,10% | 53,45% | 34,48% | 34,48% | 96,55% | 5,05 |
| INGENIERÍA ELÉCTRICA, ELECTRÓNICA Y DE COM | 137 | 58,39% | 70,07% | 54,01% | 56,20% | 57,66% | 39,42% | 28,47% | 24,82% | 29,20% | 94,16% | 4,18 |
| ECONOMÍA | 58 | 63,79% | 56,90% | 67,24% | 51,72% | 55,17% | 25,86% | 27,59% | 18,97% | 24,14% | 93,10% | 3,91 |
| CIENCIAS | 96 | 45,83% | 61,46% | 43,75% | 51,04% | 55,21% | 50,00% | 36,46% | 16,67% | 13,54% | 92,71% | 3,74 |
| ESTADÍSTICA, INFORMÁTICA Y MATEMÁTICAS | 121 | 43,80% | 61,98% | 47,93% | 50,41% | 50,41% | 28,93% | 32,23% | 18,18% | 25,62% | 84,30% | 3,60 |
| AGRONOMÍA, BIOTECNOLOGÍA Y ALIMENTACIÓN | 76 | 44,74% | 61,84% | 42,11% | 44,74% | 51,32% | 57,89% | 32,89% | 7,89% | 11,84% | 100,00% | 3,55 |
| SOCIOLOGÍA Y TRABAJO SOCIAL | 92 | 39,13% | 35,87% | 48,91% | 47,83% | 29,35% | 30,43% | 13,04% | 39,13% | 20,65% | 82,61% | 3,04 |
| INGENIERÍA | 121 | 52,07% | 52,07% | 35,54% | 36,36% | 41,32% | 36,36% | 14,88% | 9,92% | 8,26% | 86,78% | 2,87 |
| CIENCIAS HUMANAS Y DE LA EDUCACIÓN | 86 | 34,88% | 26,74% | 58,14% | 32,56% | 31,40% | 26,74% | 9,30% | 34,88% | 18,60% | 89,53% | 2,73 |
| INSTITUTO AGROBIOTECNOLOGIA | 25 | 60,00% | 36,00% | 24,00% | 40,00% | 20,00% | 44,00% | 0,00% | 16,00% | 0,00% | 92,00% | 2,40 |
| CIENCIAS DE LA SALUD | 139 | 49,64% | 33,09% | 25,18% | 31,65% | 30,94% | 26,62% | 14,39% | 14,39% | 9,35% | 79,14% | 2,35 |
| DERECHO | 72 | 56,94% | 8,33% | 40,28% | 27,78% | 16,67% | 13,89% | 5,56% | 18,06% | 6,94% | 83,33% | 1,94 |
| **Total general** | **1081** | **49,58%** | **49,21%** | **44,87%** | **44,31%** | **44,31%** | **34,60%** | **22,85%** | **20,72%** | **17,58%** | **88,53%** | **3,28** |

*Tabla 21 - % por departamentos de todos los perfiles*

El Departamento de Gestión de Empresas no solo es el que más perfiles por investigador tiene, también es el segundo con más PDI con al menos un perfil de los estudiados y también suele ocupar las primeras posiciones en el uso de cada perfil, sin fallar en ninguno. Algo parecido, en menor escala se puede decir del Departamento de Ingeniería Eléctrica. En el extremo opuesto, están los departamentos de Derecho y Ciencias de la Salud que, no solo son los que menos investigadores tienen adscritos al menos a un servicio y menos perfiles por investigador tienen, sino que también suelen ocupar las últimas posiciones en el uso de cada servicio en concreto, salvo el caso de Linkedin para Derecho.

Un aspecto a considerar en los departamentos es el diferente peso que tienen los doctorandos en su plantilla. Esto explica, al menos, algunas de las primeras y últimas posiciones. El departamento de Ciencias de la Salud tiene un 41,73% de su plantilla compuesta de doctorandos, que como hemos ido comprobando, es el grupo de menor vocación de adscripción. El departamento de Gestión de Empresas, por el contrario, solo tiene un 3,45% de doctorandos en su plantilla, lo que parece ayudar en algo a esa primera plaza que ocupa.

Tampoco la composición contractual de la plantilla explica todas las diferencias entre departamentos. Por ejemplo, el departamento de derecho, con un 16,67% de doctorandos en su plantilla, ocupa las últimas posiciones en la mayoría de indicadores. El departamento de Sociología, por el contrario, con un 22, 83% de doctorandos, suele andar por las zonas medias.

| Etiquetas de fila | % >0 | Perfiles/n | % doctorandos |
|---|---|---|---|
| CIENCIAS DE LA SALUD | 79,14% | 2,35 | 41,73% |
| SOCIOLOGÍA Y TRABAJO SOCIAL | 82,61% | 3,04 | 22,83% |
| CIENCIAS HUMANAS Y DE LA EDUCACIÒN | 89,53% | 2,73 | 18,60% |
| DERECHO | 83,33% | 1,94 | 16,67% |
| INGENIERÍA | 86,78% | 2,87 | 14,05% |
| ECONOMÍA | 93,10% | 3,91 | 12,07% |
| INSTITUTO AGROBIOTECNOLOGIA | 92,00% | 2,40 | 12,00% |
| INGENIERÍA ELÉCTRICA, ELECTRÓNICA Y DE COM | 94,16% | 4,18 | 8,76% |
| ESTADÍSTICA, INFORMÁTICA Y MATEMÁTICAS | 84,30% | 3,60 | 8,26% |
| AGRONOMÍA, BIOTECNOLOGÍA Y ALIMENTACIÓN | 100,00% | 3,55 | 6,58% |
| CIENCIAS | 92,71% | 3,74 | 6,25% |
| GESTIÓN DE EMPRESAS | 96,55% | 5,05 | 3,45% |
| Total general | 88,53% | 3,28 | 15,63% |

En la Tabla 22 podemos ver la proporción de doctorandos entre el personal de cada departamento comparado con el número de investigadores con al menos un perfil y la media de perfiles por investigador.

*Tabla 22 - % de doctorados por departamento*

En la Tabla 23 podemos ver una distribución por departamentos tomando solo las plazas ocupadas por doctores, para que la situación contractual y los diferentes pesos de algunas figuras en la composición de las plantillas no desvirtúen los resultados.

| Situación | n | % Scopus | % ORCID | % A-e | % RG | % Linkedin | % ResID | Mendeley | % GSC | Acad.edu | % >0 | Perfiles / n |
|---|---|---|---|---|---|---|---|---|---|---|---|---|
| INGENIERÍA ELÉCTRICA, ELECTRÓNICA Y DE COM | 69 | 88,41% | 78,26% | 78,26% | 71,01% | 53,62% | 50,72% | 49,28% | 47,83% | 34,78% | 95,65% | 5,52 |
| GESTIÓN DE EMPRESAS | 49 | 77,55% | 71,43% | 65,31% | 85,71% | 61,22% | 59,18% | 36,73% | 38,78% | 38,78% | 95,92% | 5,35 |
| ECONOMÍA | 36 | 75,00% | 66,67% | 91,67% | 66,67% | 63,89% | 33,33% | 22,22% | 33,33% | 22,22% | 100,00% | 4,75 |
| ESTADÍSTICA, INFORMÁTICA Y MATEMÁTICAS | 70 | 81,43% | 72,86% | 62,86% | 67,14% | 44,29% | 54,29% | 30,00% | 32,86% | 22,86% | 94,29% | 4,69 |
| SOCIOLOGÍA Y TRABAJO SOCIAL | 39 | 61,54% | 74,36% | 82,05% | 51,28% | 51,28% | 23,08% | 25,64% | 35,90% | 46,15% | 100,00% | 4,51 |
| AGRONOMÍA, BIOTECNOLOGÍA Y ALIMENTACIÓN | 46 | 76,09% | 67,39% | 58,70% | 67,39% | 30,43% | 52,17% | 52,17% | 19,57% | 10,87% | 100,00% | 4,35 |
| CIENCIAS | 58 | 77,59% | 68,97% | 56,90% | 60,34% | 46,55% | 44,83% | 46,55% | 15,52% | 17,24% | 96,55% | 4,34 |
| CIENCIAS DE LA SALUD | 39 | 74,36% | 61,54% | 51,28% | 56,41% | 48,72% | 38,46% | 35,90% | 25,64% | 33,33% | 97,44% | 4,26 |
| INGENIERÍA | 60 | 70,00% | 56,67% | 51,67% | 51,67% | 55,00% | 28,33% | 41,67% | 13,33% | 8,33% | 95,00% | 3,77 |
| CIENCIAS HUMANAS Y DE LA EDUCACIÓN | 30 | 43,33% | 53,33% | 86,67% | 33,33% | 43,33% | 16,67% | 23,33% | 26,67% | 43,33% | 96,67% | 3,70 |
| INSTITUTO AGROBIOTECNOLOGIA | 15 | 40,00% | 60,00% | 26,67% | 26,67% | 60,00% | 0,00% | 33,33% | 0,00% | 20,00% | 93,33% | 2,67 |
| DERECHO | 38 | 10,53% | 42,11% | 57,89% | 15,79% | 57,89% | 7,89% | 15,79% | 10,53% | 13,16% | 94,74% | 2,32 |
| Total general | 549 | 69,40% | 66,12% | 65,21% | 58,47% | 50,64% | 38,80% | 36,25% | 27,14% | 25,32% | 96,54% | 4,37 |

*Tabla 23 - % por departamentos con PDI doctor*

Se puede apreciar como el departamento de Ciencias de la Salud sube puestos una vez filtrado ese alto porcentaje de doctorandos. Gestión de Empresas sigue ocupando posiciones altas, a pesar de verse despojado de la ventaja que suponía tener una bajo porcentaje de doctorandos.

El departamento de Ciencias sí que baja posiciones, porque se beneficiaba de su bajo porcentaje de doctorandos, y Sociología sube un poco sin el lastre del elevado número de doctorados.

También vemos que Linkedin deja de ser el perfil con más adscritos, siendo superado por Scopus, Orcid, Academica-e y ResearchGate.

## 4.3 SITUACIÓN CONTRACTUAL

En la *Tabla 24* se muestra el cruce comparativo de las situaciones contractuales[15] con los distintos perfiles. El color de las celdas, de verde a rojo, marca el valor del dato entre todos los perfiles. Los puntos de color, de verde a negro, marcan el valor dentro del mismo perfil.

El orden de la situación contractual lo marca el dato de la media de perfiles por investigdor (perfiles/n).

Atendiendo a la situación contractual, el dato de investigadores con al menos un perfil de los estudiados (% >0) y el de número de perfiles por investigador (Perfiles/n), van muy parejos, al contrario de lo que sucedía al comparar los departamentos. En efecto, contratados doctores, catedráticos, ayudantes doctores y titulares ocupan los primeros lugares, con bastante ventaja sobre el resto de categorías contractuales. Investigadores doctores y becarios están en una zona media, y eméritos, asociados, investigadores no doctores y doctorandos ocupan las últimas posiciones.

| Situación | n | % Link | % Scopus | % A-e | % ORCID | % RG | % Mend | % ResID | Acad.edu | % GSC | % >0 | Perfiles/n |
|---|---|---|---|---|---|---|---|---|---|---|---|---|
| Contratado doctor | 126 | 47,62% | 69,05% | 72,22% | 72,22% | 65,87% | 46,03% | 44,44% | 35,71% | 37,30% | 99,21% | 4,90 |
| Catedrático | 83 | 49,40% | 75,90% | 80,72% | 63,86% | 53,01% | 28,92% | 45,78% | 25,30% | 27,71% | 97,59% | 4,51 |
| Ayudante doctor | 54 | 57,41% | 61,11% | 55,56% | 77,78% | 64,81% | 40,74% | 31,48% | 25,93% | 35,19% | 96,30% | 4,50 |
| Titular | 246 | 50,81% | 71,95% | 63,01% | 61,79% | 56,10% | 33,33% | 38,62% | 21,95% | 23,17% | 95,12% | 4,21 |
| Investigador doctor | 40 | 52,50% | 52,50% | 37,50% | 62,50% | 52,50% | 32,50% | 17,50% | 12,50% | 7,50% | 95,00% | 3,28 |
| Becario | 110 | 58,18% | 40,91% | 26,36% | 36,36% | 50,91% | 65,45% | 12,73% | 17,27% | 14,55% | 94,55% | 3,23 |
| Emérito | 16 | 43,75% | 50,00% | 75,00% | 18,75% | 18,75% | 6,25% | 25,00% | 6,25% | 87,50% | 2,50 | |
| Asociado | 155 | 48,39% | 33,55% | 41,94% | 28,39% | 33,55% | 25,16% | 5,81% | 20,00% | 9,68% | 85,81% | 2,46 |
| Investigador | 82 | 60,98% | 35,37% | 20,73% | 14,63% | 36,59% | 35,37% | 8,54% | 14,63% | 6,10% | 84,15% | 2,33 |
| Doctorando | 169 | 36,69% | 10,06% | 2,37% | 10,06% | 10,06% | 20,12% | 1,78% | 11,24% | 2,37% | 63,31% | 1,05 |
| **Total general** | **1081** | **49,58%** | **49,21%** | **44,87%** | **44,31%** | **44,31%** | **34,60%** | **22,85%** | **20,72%** | **17,58%** | **88,53%** | **3,28** |
*Tabla 24 - % por situación contractual de todos los perfiles*

Observamos un lógico aumento del número de perfiles según esté más avanzada la carrera investigadora en la mayoría de perfiles estudiados. La excepción es Linkedin y, quizá también, Mendeley, debido a que son servicios cuya principal vocación no es el mantenimiento de perfiles de investigación.

Un dato fuera de lo previsible es el elevado porcentaje de investigadores no doctores y becarios que son usuarios de Linkedin. Esto quizá se explique por ser investigadores más jóvenes y, por tanto, con más contacto con redes sociales en general, además, Linkedin es una herramienta muy utilizada para la búsqueda de empleo, algo que está dentro de los intereses de este grupo. Otro dato que puede llamar la atención es que el uso de Mendeley lo lidera el grupo de los becarios. La explicación puede ser que, en los grupos de investigación, se suele encargar a los becarios el trabajo de redactar la bibliografía y elaborar las citas y referencias; también son los encargados de formarse para luego repercutir esa formación en el resto del grupo. Por estos motivos, están más familiarizados con Mendeley.

---

[15] Becario engloba todos los contratos predoctorales; Asociado engloba: asociado 1, asociado 2 y asociado 3; Investigador engloba: ayudante de proyecto y colaborador de proyecto; Investigador doctor engloba: colaborador doctor de proyecto y otros contratos postdoctorales.

## 4.4 Grupo de investigación

Los grupos de investigación de la UPNA tienen un número variable de miembros, desde grupos como Pensamiento Político o Derecho Internacional Público, con un solo miembro, al grupo Comunicaciones Ópticas, con 55 miembros.

| Miembros | n | % Link | % Scopus | % A-e | % ORCID | % RG | % Mend | % ResID | % Acad.edu | % GSC | % >0 | Perfiles / n |
|---|---|---|---|---|---|---|---|---|---|---|---|---|
| 0-5 | 106 | 44,34% | 56,60% | 51,89% | 45,28% | 37,74% | 29,25% | 23,58% | 17,92% | 12,26% | 93,40% | 3,19 |
| 6-10 | 223 | 40,36% | 52,47% | 49,78% | 43,95% | 46,19% | 31,39% | 25,56% | 23,77% | 18,83% | 86,55% | 3,32 |
| 11-15 | 279 | 54,84% | 47,67% | 40,50% | 44,44% | 46,95% | 34,41% | 26,16% | 19,00% | 16,49% | 89,25% | 3,30 |
| 16-20 | 178 | 47,75% | 43,26% | 44,94% | 46,63% | 42,70% | 27,53% | 19,66% | 26,97% | 22,47% | 83,71% | 3,22 |
| >20 | 295 | 54,58% | 49,15% | 42,71% | 42,71% | 43,73% | 43,39% | 19,32% | 17,29% | 16,61% | 90,51% | 3,29 |

*Tabla 25 - % por tamaño de grupo*

Según apreciamos en la Tabla 25, no parece haber ningún patrón determinado por el tamaño del grupo para una mayor o menos adscripción a los servicios de perfiles. En Scopus, Academica-e, ResearcherID y número de investigadores con al menos un perfil (%>0), parece que, cuanto más pequeño es el grupo, mayor es la adscripción, disminuyendo cuando aumenta el tamaño del grupo. En Linkedin, Mendeley, GSC y media de perfiles por investigador (Perfiles/n), parece ser lo contrario: cuanto mayor es el grupo, mayor es la adscripción. Por fin, en Orcid, ResearchGate y Academia.edu, los datos son erráticos y tan pronto suben como bajan según aumenta el tamaño del grupo de investigación.

En la *Tabla 26* se muestra el porcentaje de adscripción a los distintos perfiles por grupo de investigación. El color de las celdas, de verde a rojo, marca el valor del dato entre todos los perfiles. Los puntos de color, de verde a negro, marcan el valor dentro del mismo perfil.

El orden de los grupos de investigación lo marca el dato de la media de perfiles por investigador (perfiles/n).

En general, observamos una fuerte correlación entre la distribución de los perfiles por grupos de investigación y por departamentos. En efecto, los grupos de investigación de departamentos como Gestión de Empresas, Economía o Ingeniería Eléctrica son proclives a tener perfiles, y los grupos de los departamentos de Derecho y Ciencias de la Salud son lo que menos perfiles tienen, repitiendo la situación que se ve en los departamentos.

Al igual que con los departamentos también los grupos de investigación se ven muy afectados por su composición, atendiendo a la situación contractual. Algunos de los que ocupan los últimos lugares, tales como el grupo Patología Crónica (55,5%), Estudios de Derecho Público (40%), Cultura y Desarrollo (45,45%) y Fisiopatología (94,4%), pagan un alto porcentaje de doctorandos en su composición.

Otros como el grupo Economía Agraria (0%), Grupo de Antenas (7,14%), Grupo de Redes (0%), Genética y Microbiología (0%) o Grupo de Marketing (8,33%) se benefician del bajo número de doctorandos en su plantilla.

| Grupo | n | % Linkedin | % Scopus | % A-e | % ORCID | % RG | % Mendeley | % ResID | % Acad.edu | % GSC | % >0 | Perfiles / n |
|---|---|---|---|---|---|---|---|---|---|---|---|---|
| Economía agraria | 6 | 100,00% | 66,67% | 100,00% | 83,33% | 66,67% | 33,33% | 66,67% | 0,00% | 50,00% | 100,00% | 5,67 |
| Grupo de Antenas | 14 | 78,57% | 64,29% | 71,43% | 64,29% | 57,14% | 78,57% | 50,00% | 35,71% | 35,71% | 100,00% | 5,36 |
| Grupo de redes, sistemas y servic | 6 | 83,33% | 83,33% | 83,33% | 50,00% | 83,33% | 33,33% | 50,00% | 16,67% | 50,00% | 100,00% | 5,33 |
| Genética y microbiología | 6 | 50,00% | 83,33% | 83,33% | 66,67% | 83,33% | 33,33% | 83,33% | 16,67% | 33,33% | 100,00% | 5,33 |
| Grupo marketing | 12 | 50,00% | 66,67% | 50,00% | 66,67% | 91,67% | 75,00% | 50,00% | 33,33% | 33,33% | 91,67% | 5,17 |
| Economía de la Empresa | 13 | 46,15% | 92,31% | 53,85% | 76,92% | 92,31% | 30,77% | 38,46% | 38,46% | 46,15% | 100,00% | 5,15 |
| Psicología Clínica y Psicopatología | 11 | 45,45% | 63,64% | 63,64% | 63,64% | 63,64% | 45,45% | 72,73% | 45,45% | 45,45% | 90,91% | 5,09 |
| Mercados financieros | 8 | 25,00% | 62,50% | 87,50% | 62,50% | 87,50% | 25,00% | 100,00% | 50,00% | 0,00% | 100,00% | 5,00 |
| Inteligencia artificial y razonamie | 20 | 50,00% | 85,00% | 65,00% | 60,00% | 65,00% | 30,00% | 50,00% | 20,00% | 70,00% | 95,00% | 4,95 |
| Organización de empresas | 16 | 75,00% | 75,00% | 43,75% | 56,25% | 81,25% | 56,25% | 37,50% | 31,25% | 31,25% | 93,75% | 4,88 |
| Ecología y medio ambiente | 9 | 22,22% | 100,00% | 66,67% | 55,56% | 77,78% | 66,67% | 33,33% | 11,11% | 33,33% | 100,00% | 4,67 |
| Social Equilibrium and Economic D | 6 | 66,67% | 66,67% | 83,33% | 66,67% | 66,67% | 16,67% | 16,67% | 33,33% | 33,33% | 83,33% | 4,50 |
| Economía pública y regional | 8 | 62,50% | 87,50% | 75,00% | 50,00% | 50,00% | 12,50% | 62,50% | 25,00% | 25,00% | 87,50% | 4,50 |
| Comunicaciones ópticas y aplicaci | 55 | 70,91% | 65,45% | 54,55% | 56,36% | 58,18% | 47,27% | 34,55% | 21,82% | 36,36% | 94,55% | 4,45 |
| Investigación en contabilidad | 6 | 50,00% | 66,67% | 16,67% | 66,67% | 100,00% | 0,00% | 50,00% | 50,00% | 33,33% | 100,00% | 4,33 |
| Sociología rural, movilidad e inve | 11 | 63,64% | 63,64% | 63,64% | 81,82% | 45,45% | 27,27% | 18,18% | 45,45% | 18,18% | 100,00% | 4,27 |
| Fruticultura y viticultura avanzada | 9 | 77,78% | 55,56% | 33,33% | 55,56% | 44,44% | 66,67% | 33,33% | 0,00% | 22,22% | 100,00% | 3,89 |
| Applied linguistics: Language Acq | 7 | 28,57% | 57,14% | 42,86% | 42,86% | 71,43% | 42,86% | 14,29% | 57,14% | 28,57% | 100,00% | 3,86 |
| Problemas diferenciales y aproxim | 7 | 14,29% | 100,00% | 42,86% | 57,14% | 57,14% | 42,86% | 42,86% | 14,29% | 14,29% | 100,00% | 3,86 |
| Producción animal y calidad tec | 13 | 30,77% | 76,92% | 53,85% | 69,23% | 61,54% | 46,15% | 46,15% | 0,00% | 0,00% | 100,00% | 3,85 |
| DECYL (Datos, Estadística, Calidad | 19 | 57,89% | 68,42% | 26,32% | 57,89% | 52,63% | 42,11% | 26,32% | 26,32% | 26,32% | 78,95% | 3,84 |
| Análisis Económico | 11 | 63,64% | 45,45% | 54,55% | 45,45% | 63,64% | 36,36% | 27,27% | 18,18% | 27,27% | 90,91% | 3,82 |
| Fisiología vegetal y Agrobiología | 22 | 59,09% | 45,45% | 40,91% | 45,45% | 63,64% | 63,64% | 40,91% | 13,64% | 9,09% | 95,45% | 3,82 |
| Comunicación, señales y microon | 16 | 37,50% | 75,00% | 50,00% | 75,00% | 62,50% | 12,50% | 31,25% | 31,25% | 6,25% | 87,50% | 3,81 |
| Ingeniería biomédica | 15 | 60,00% | 80,00% | 53,33% | 60,00% | 46,67% | 13,33% | 26,67% | 26,67% | 13,33% | 100,00% | 3,80 |
| THERRAE: Teledetección, Hidrolo | 21 | 52,38% | 57,14% | 47,62% | 52,38% | 57,14% | 57,14% | 19,05% | 9,52% | 23,81% | 95,24% | 3,76 |
| José María Lacarra | 7 | 71,43% | 14,29% | 57,14% | 42,86% | 42,86% | 42,86% | 14,29% | 57,14% | 28,57% | 100,00% | 3,71 |
| Mecatrónica agraria | 7 | 42,86% | 85,71% | 57,14% | 57,14% | 42,86% | 42,86% | 28,57% | 0,00% | 14,29% | 100,00% | 3,71 |
| GIAS (Grupo de Investigación en A | 9 | 44,44% | 33,33% | 55,56% | 44,44% | 22,22% | 44,44% | 33,33% | 55,56% | 33,33% | 77,78% | 3,67 |
| Ingeniería Eléctrica, Electrónica d | 23 | 26,09% | 69,57% | 52,17% | 52,17% | 60,87% | 43,48% | 4,35% | 21,74% | 34,78% | 86,96% | 3,65 |
| Estadística espacial | 11 | 54,55% | 45,45% | 63,64% | 36,36% | 45,45% | 27,27% | 54,55% | 18,18% | 18,18% | 81,82% | 3,64 |
| Adquisición de conocimiento y m | 9 | 22,22% | 77,78% | 33,33% | 66,67% | 66,67% | 33,33% | 33,33% | 11,11% | 11,11% | 100,00% | 3,56 |
| Protección de cultivos (PC) | 22 | 54,55% | 40,91% | 45,45% | 40,91% | 45,45% | 68,18% | 27,27% | 22,73% | 9,09% | 100,00% | 3,55 |
| ALTER. Grupo de investigación | 19 | 36,84% | 36,84% | 52,63% | 52,63% | 47,37% | 52,63% | 15,79% | 31,58% | 26,32% | 94,74% | 3,53 |
| Organización industrial, territorio | 6 | 66,67% | 83,33% | 100,00% | 33,33% | 50,00% | 0,00% | 0,00% | 0,00% | 16,67% | 100,00% | 3,50 |
| Álgebra. Aplicaciones. | 12 | 50,00% | 50,00% | 66,67% | 33,33% | 58,33% | 25,00% | 25,00% | 25,00% | 16,67% | 91,67% | 3,50 |
| Reactores Químicos y Procesos pa | 13 | 61,54% | 46,15% | 15,38% | 38,46% | 53,85% | 53,85% | 30,77% | 15,38% | 7,69% | 84,62% | 3,23 |
| Proyectos, ingeniería rural y ener | 13 | 38,46% | 46,15% | 15,38% | 53,85% | 53,85% | 69,23% | 38,46% | 0,00% | 7,69% | 84,62% | 3,23 |
| Efimec: Ética, filosofía y metodolo | 9 | 44,44% | 55,56% | 22,22% | 55,56% | 33,33% | 44,44% | 11,11% | 33,33% | 22,22% | 77,78% | 3,22 |
| Hizkuntzalaritzako ikerketa/Inves | 16 | 50,00% | 25,00% | 75,00% | 31,25% | 37,50% | 37,50% | 12,50% | 31,25% | 18,75% | 93,75% | 3,19 |
| Historia y Economía | 26 | 34,62% | 30,77% | 65,38% | 42,31% | 34,62% | 38,46% | 23,08% | 26,92% | 19,23% | 96,15% | 3,15 |
| Propiedades físicas y aplicaciones | 11 | 54,55% | 45,45% | 18,18% | 54,55% | 54,55% | 27,27% | 18,18% | 9,09% | 27,27% | 81,82% | 3,09 |
| Biofilms microbianos | 12 | 58,33% | 41,67% | 50,00% | 25,00% | 41,67% | 16,67% | 33,33% | 16,67% | 16,67% | 91,67% | 3,00 |
| Gestión y manejo sostenible de s | 7 | 28,57% | 71,43% | 42,86% | 28,57% | 28,57% | 42,86% | 28,57% | 28,57% | 0,00% | 85,71% | 3,00 |
| Diseño industrial | 10 | 50,00% | 70,00% | 50,00% | 40,00% | 30,00% | 30,00% | 10,00% | 20,00% | 0,00% | 80,00% | 3,00 |
| Economía de la salud | 15 | 73,33% | 33,33% | 33,33% | 40,00% | 46,67% | 26,67% | 13,33% | 13,33% | 13,33% | 93,33% | 2,93 |
| Evaluación y desarrollo de la perc | 10 | 20,00% | 10,00% | 40,00% | 50,00% | 60,00% | 30,00% | 0,00% | 40,00% | 40,00% | 70,00% | 2,90 |
| Matemáticas del orden | 9 | 33,33% | 55,56% | 66,67% | 44,44% | 22,22% | 11,11% | 11,11% | 22,22% | 22,22% | 77,78% | 2,89 |
| Agrobiotecnología | 30 | 50,00% | 43,33% | 30,00% | 46,67% | 26,67% | 50,00% | 10,00% | 13,33% | 3,33% | 93,33% | 2,73 |
| Educación, desarrollo profesional | 7 | 14,29% | 28,57% | 57,14% | 28,57% | 28,57% | 14,29% | 14,29% | 71,43% | 14,29% | 100,00% | 2,71 |
| Acústica | 7 | 14,29% | 71,43% | 42,86% | 28,57% | 71,43% | 28,57% | 14,29% | 0,00% | 0,00% | 85,71% | 2,71 |
| Grupo de investigación en Sabere | 21 | 52,38% | 38,10% | 28,57% | 42,86% | 28,57% | 52,38% | 4,76% | 9,52% | 14,29% | 95,24% | 2,71 |
| Sistemas distribuidos | 24 | 54,17% | 54,17% | 33,33% | 37,50% | 37,50% | 16,67% | 20,83% | 8,33% | 8,33% | 75,00% | 2,71 |
| Cambios sociales | 20 | 30,00% | 30,00% | 45,00% | 50,00% | 25,00% | 10,00% | 10,00% | 50,00% | 20,00% | 75,00% | 2,70 |
| Ingeniería de materiales y fabrica | 21 | 61,90% | 57,14% | 57,14% | 23,81% | 23,81% | 23,81% | 0,00% | 9,52% | 0,00% | 90,48% | 2,57 |
| Ingeniería Térmica y de Fluidos | 15 | 66,67% | 20,00% | 40,00% | 26,67% | 40,00% | 40,00% | 13,33% | 6,67% | 0,00% | 80,00% | 2,53 |
| Administración Pública | 11 | 63,64% | 9,09% | 45,45% | 36,36% | 9,09% | 18,18% | 0,00% | 27,27% | 18,18% | 81,82% | 2,27 |
| Derecho privado | 16 | 56,25% | 12,50% | 43,75% | 31,25% | 37,50% | 6,25% | 12,50% | 12,50% | 12,50% | 87,50% | 2,25 |
| Hugo Grocio | 18 | 55,56% | 11,11% | 50,00% | 33,33% | 5,56% | 22,22% | 0,00% | 33,33% | 5,56% | 88,89% | 2,17 |
| Control de la expresión génica | 12 | 41,67% | 41,67% | 16,67% | 41,67% | 33,33% | 33,33% | 0,00% | 8,33% | 0,00% | 91,67% | 2,17 |
| Tecnología, control y seguridad al | 13 | 46,15% | 53,85% | 15,38% | 7,69% | 38,46% | 38,46% | 7,69% | 7,69% | 0,00% | 84,62% | 2,15 |
| Desarrollo psicológico y educació | 7 | 42,86% | 42,86% | 71,43% | 14,29% | 14,29% | 14,29% | 0,00% | 14,29% | 0,00% | 85,71% | 2,14 |
| Ejercicio Físico, Salud y Calidad de | 30 | 63,33% | 26,67% | 10,00% | 16,67% | 33,33% | 20,00% | 10,00% | 23,33% | 3,33% | 73,33% | 2,07 |
| Tecnología Energética | 6 | 0,00% | 33,33% | 0,00% | 33,33% | 33,33% | 50,00% | 16,67% | 16,67% | 16,67% | 66,67% | 2,00 |
| Ingeniería mecánica aplicada y co | 15 | 60,00% | 33,33% | 6,67% | 20,00% | 26,67% | 6,67% | 13,33% | 6,67% | 6,67% | 93,33% | 1,80 |
| Epidemiología | 15 | 46,67% | 26,67% | 26,67% | 33,33% | 13,33% | 6,67% | 6,67% | 6,67% | 13,33% | 80,00% | 1,80 |
| Didáctica de la Matemática | 9 | 22,22% | 0,00% | 22,22% | 44,44% | 22,22% | 22,22% | 11,11% | 11,11% | 22,22% | 66,67% | 1,78 |
| Derecho del trabajo | 7 | 42,86% | 0,00% | 42,86% | 0,00% | 28,57% | 14,29% | 0,00% | 14,29% | 0,00% | 71,43% | 1,43 |
| Cultura y Desarrollo Lera-ikergune | 11 | 45,45% | 0,00% | 27,27% | 9,09% | 0,00% | 18,18% | 0,00% | 27,27% | 9,09% | 63,64% | 1,36 |
| Estudios de Derecho Público sobr | 10 | 60,00% | 10,00% | 10,00% | 20,00% | 10,00% | 20,00% | 0,00% | 10,00% | 0,00% | 70,00% | 1,30 |
| Fisiopatología y práctica clínica | 18 | 33,33% | 11,11% | 0,00% | 16,67% | 16,67% | 5,56% | 0,00% | 0,00% | 0,00% | 44,44% | 0,83 |
| Patología crónica y del envejecim | 9 | 11,11% | 0,00% | 11,11% | 0,00% | 0,00% | 33,33% | 0,00% | 11,11% | 0,00% | 55,56% | 0,67 |
| **Total general** | **975** | **50,15%** | **48,41%** | **44,10%** | **44,21%** | **45,03%** | **35,18%** | **22,77%** | **21,03%** | **18,15%** | **88,00%** | **3,29** |

*Tabla 26 - % por grupo de investigación con más de 5 miembros*

## 4.5 Género

En la *Tabla 27* se muestra el porcentaje por género en los distintos perfiles. El color de las celdas, de verde a rojo, marca el valor del dato entre todos los perfiles. Los puntos de color, de verde a negro, marcan el valor dentro del mismo perfil.

Según observamos, suele haber un porcentaje mayor de hombres que de mujeres adscritos a los diferentes perfiles. Las excepciones son Mendeley (H 31,32% - M 38,89%), Linkedin (H 48,78% - M 50,64%) y, en una pequeña proporción Academia.edu (H 20,55% - M 20,94%).

| Género | n | % Linkedin | % Scopus | % A-e | % ORCID | % RG | % Mendeley | % ResID | % Acad.edu | % GSC | % >0 | Perfiles / n |
|---|---|---|---|---|---|---|---|---|---|---|---|---|
| Hombre | 613 | 48,78% | 54,00% | 49,27% | 47,80% | 46,66% | 31,32% | 25,94% | 20,55% | 20,23% | 87,77% | 3,45 |
| Mujer | 468 | 50,64% | 42,95% | 39,10% | 39,74% | 41,24% | 38,89% | 18,80% | 20,94% | 14,10% | 89,53% | 3,06 |
| Total general | 1081 | 49,58% | 49,21% | 44,87% | 44,31% | 44,31% | 34,60% | 22,85% | 20,72% | 17,58% | 88,53% | 3,28 |

*Tabla 27 - % por género*

Estas diferencias no se explican examinando el género por departamento o por situación contractual, ya que, salvo algunas excepciones anecdóticas, se repite este modelo para todos los departamentos y grupos contractuales.

El número de mujeres con al menos un perfil supera casi en dos puntos al de hombres (H 87,77% - M 89,53%), sin embargo, la media de perfiles por persona es levemente mayor en el grupo de hombres que en el de mujeres (H 3,45 – M 3,06)

La conclusión general, que tampoco tiene por qué ser universal, dadas las pequeñas diferencias existentes, es que las mujeres son más proclives a tener al menos un perfil, mientras que los hombres, de media, están adscritos a más perfiles.

Los perfiles en Linkedin y Mendeley, según hemos visto anteriormente, ofrecen distintos comportamientos que el resto. Esto se apreciaba en el análisis por situación contractual y en el análisis por departamentos. En el análisis por género, también observamos un comportamiento distinto, lo que parece indicar que, en la práctica, no son percibidos, o siguen distintas pautas que el resto de servicios de perfiles.

# 5 Conclusiones

En cuanto a la efectividad de las actividades de formación y asistencia llevadas a cabo desde la biblioteca para potenciar la adscripción a los perfiles de investigación, podríamos estar relativamente satisfechos: un 49,21% tiene perfil en Scopus (69,4% en el grupo de doctores[16]), un 44,31% en Orcid (66,12%), un 34,6% en Mendeley (36,25%), y un 44,87% ha depositado algo en Academica-e (65,21%). Menos satisfechos tenemos que estar con un 22,85 que disponen de un ResearcherID (38,8%) o del escaso 17,58% con perfil en Google Scholar Citations (27,14%).

En los resultados de las segmentaciones, observamos que apenas hay diferencias por género a la hora de adscribirse a los diferentes perfiles, aunque sí se aprecia una ligera ventaja de los hombres respecto de las mujeres; que hay departamentos más proclives que otros a la hora de apuntarse en perfiles de investigación y, sobre todo, se observa un fuerte condicionamiento según la situación contractual o etapa de la carrera investigadora en que se encuentre: catedráticos, titulares y contratados doctores se inscriben en los perfiles con mucha más frecuencia que asociados y doctorandos.

Un dato para la reflexión es que, de los perfiles de personal investigador de la UPNA existentes en ORCID, una gran parte no incluye ningún dato bibliográfico ni remite a otro sistema de perfil investigador. Efectivamente, hay 489 (44,31%) perfiles de la UPNA en Orcid, pero solamente 297 (27,47%) tienen algún dato bibliográfico o de remisión a otro servicio. No sabemos cuán preocupante puede ser esta ausencia de datos. El investigador o investigadora obtiene su perfil Orcid para responder a la necesidad inmediata de disponer del identificador, porque se lo han pedido para algún trámite. La urgencia está solucionada, pero el perfil Orcid permanecerá vacío y si alguien va a consultarlo, lo encontrará vacío.

Otro problema que detectamos es la baja afiliación a Google Scholar Citations, especialmente lamentable en los departamentos de humanidades y ciencias sociales. En estas áreas del conocimiento –donde abundan las revistas de ámbito local y se publica y difunde en formato libro– es complicado obtener datos del impacto de las publicaciones, y GSC sería de gran ayuda, puesto que está demostrado que aporta un mayor número de citas en todas las disciplinas, y especialmente en humanidades y ciencias sociales (Martín-Martín et al., 2018).

---

[16] Consideramos aquí las situaciones contractuales de catedrático, titular, contratado doctor, investigador doctor y ayudante doctor.

# 6 Referencias

https://doi.org/10.5860/crl.78.2.171

VERHAAR, PETER. (2018). *ORCID implementation at Leiden University [Presentación]*. https://bit.ly/2JNgkCI